\documentclass[aps,twocolumn,showpacs,superscriptaddress,amsmath,amssymb,pre,nofootinbib]{revtex4-1}
\usepackage{epsf,amsmath,amssymb,amsfonts,verbatim,color,multirow,pifont}
\usepackage{graphicx}
\usepackage{dcolumn}
\usepackage{bm}
\usepackage{txfonts}
\usepackage{hyperref}
\usepackage{soul}

\begin{document}
\title{Thermodynamic uncertainty relation for underdamped Langevin systems driven by a velocity-dependent force}
\author{Jae Sung Lee}
\author{Jong-Min Park}
\author{Hyunggyu~Park}
\email{hgpark@kias.re.kr}
\affiliation{School of Physics and Quantum Universe Center, Korea Institute for Advanced Study, Seoul 02455, Korea}
\newcommand{\revise}[1]{{\color{red}#1}}

\date{\today}

\begin{abstract}
Recently, it has been shown that there is a trade-off relation between thermodynamic cost and current fluctuations, referred to as the thermodynamic uncertainty relation (TUR). The TUR has been derived for various processes, such as discrete-time Markov jump processes and overdamped Langevin dynamics. For underdamped dynamics, it has recently been reported that some modification is necessary for application of the TUR. In this study, we present a more generalized TUR, applicable to a system driven by a velocity-dependent force in the context of underdamped Langevin dynamics, by extending the theory of Vu and Hasegawa [preprint arXiv:1901.05715]. We show that our TUR accurately describes the trade-off properties of a molecular refrigerator (cold damping), Brownian dynamics in a magnetic field, and an active particle system.
\end{abstract}

\pacs{05.70.Ln, 05.70.-a, 05.60.Gg}

\maketitle

\section {Introduction}

The thermodynamic uncertainty relation (TUR) is a trade-off relation between current fluctuations and entropy production (EP)~\cite{Barato,Gingrich1}. Generally, for an accumulated current $\Theta$ over a given time period $\mathcal{T}$, such as work, displacement, etc., the TUR states that the relative fluctuation multiplied by the EP is always larger than or equal to $2 k_\textrm{B}$ in the steady state; that is, the original TUR can be written as
\begin{align}
\mathcal{Q}_\textrm{ori} (\Theta) \equiv \frac{ \textrm{Var}[\Theta] }{ \langle \Theta \rangle_\textrm{s}^2 } \sigma_\textrm{s} \mathcal{T} \geq 2k_\textrm{B} , \label{eq:originalQ}
\end{align}
where $\langle \cdots \rangle_\textrm{s}$ denotes a steady-state average, $\textrm{Var}[\Theta] = \langle \Theta^2\rangle_\textrm{s} -\langle \Theta \rangle_\textrm{s}^2   $ is the variance in $\Theta$, $\sigma_\textrm{s}$ is the steady-state EP rate, and $k_\textrm{B}$ is the Boltzmann constant. Note that we use the subscript `ori' to distinguish the original TUR from other modified TURs. This relation implies that it costs a large amount of EP (heat dissipation) to achieve high accuracy (low relative fluctuation) with a stochastic motion.

The TUR was first discovered in a biological network~\cite{Barato}. It has since been derived for a continuous-time Markov jump process over a finite time~\cite{Horowitz,Gingrich2,Pietzonka1}, as well as in the long-time limit~\cite{Gingrich1}, and for an overdamped Langevin system~\cite{Gingrich2, Dechant1,Vu2}. It has also been shown that the TUR should be modified for discrete-time Markov jump processes~\cite{Proesmans}, linear-response systems~\cite{Macieszczak}, and periodically driven systems~\cite{Barato1,Barato2,Koyuk}. Moreover,
the TUR was utilized in~\cite{Pietzonka} for understanding the relations between  the power, efficiency, constancy of a heat engine~\cite{Benenti, Brandner, Allahverdyan1, Karel, Campisi, Shiraish, Holubec1, Andreas2,JSLee,JSLee1}.

The validity of the TUR was questioned recently for underdamped Langevin dynamics~\cite{FPS}. Subsequently, Vu and Hasegawa~\cite{Vu1} demonstrated that the original TUR, Eq.~\eqref{eq:originalQ}, can be violated for a squared velocity current in equilibrium and for displacement of the Brownian particle in a tilted periodic potential. They derived a modified TUR for the underdamped dynamics ~\cite{Vu1}:
\begin{align}
\mathcal{Q}_\textrm{u} (\Theta) \equiv \frac{ \textrm{Var}[\Theta] }{ \langle \Theta \rangle_\textrm{s}^2 } \Sigma_\textrm{u} \geq 2k_\textrm{B},~~~\Sigma_\textrm{u} \equiv \mathcal{T}(9 \sigma_\textrm{s} + 4 \Upsilon) +\Omega \label{eq:underdampedQ}
\end{align}
where $\Upsilon$ is the dynamical activity and $\Omega$ is a boundary term defined as in Eq.~\eqref{eq:Upsilon_Omega}.

However, Eq.~\eqref{eq:underdampedQ} is derived under the assumption that an external force is only position-dependent, and not velocity-dependent. Therefore, Eq.~\eqref{eq:underdampedQ} cannot be applied to a system such as that studied by Chun et al.~\cite{Chun}, wherein a charged Brownian particle moves under a magnetic field; specifically, they showed that Eq.~\eqref{eq:originalQ} can be violated when a magnetic field and a rotational force are applied simultaneously. To account for the effect of a Lorentz force on the TUR, a more generalized TUR is required, taking into consideration a velocity-dependent force in the underdamped dynamics. Velocity-dependent force plays a key role in many important contexts, such as molecular refrigerators (cold damping)~\cite{Liang, Cohadon, Pinard, Mertz, Jourdan, Kim}, collective motions of active/passive Brownian particles with velocity-dependent interactions~\cite{vicsek,tailleur,sevilla,noh,ams,ams2,GC}, and certain active matter dynamics~\cite{mizuno2007,cugliandolo,tailleur2015,wijland2015,Fodor,Madal}.

In this work, we extend the uncertainty relation, Eq.\eqref{eq:underdampedQ} so that it is applicable to underdamped Langevin systems with a general velocity-dependent force. We find that a velocity-dependent force only changes the dynamical activity term as presented in Eq.~\eqref{eq:Upsilon_Omega}. We examine the applicability of our inequality to three physical systems driven by velocity-dependent forces: a cold-damping problem, a magnetic-field involved problem, and an active matter problem. From these concrete examples, we show that our inequality is valid for a system driven by a velocity-dependent force, while the original TUR~\cite{Barato} and that of Vu and Hasegawa's~\cite{Vu1} do not hold. We also identify several conditions allowing the lowest bound of the inequality to be attained, which was claimed to be impossible in a previous study~\cite{Vu1}.

The paper is organized as follows. In Section 2, we explain our model system and derive the generalized TUR for the underdamped Langevin system with a velocity-dependent force. The main results are presented in Eq.~\eqref{eq:main_ineq}. In Section 3, we calculate our inequality for three cases: cold-damping, a charged particle in a magnetic field, and an active matter system. In Section 4, we conclude the paper with a brief summary and discussion.

\section {Model and generalized TUR}

We consider a $N$-particle underdamped Langevin system, where the $i$-th particle ($i=1,\cdots,N$) is in contact with a heat reservoir with temperature $T_i$. Define $x_i$ and $v_i$ as position and velocity of the $i$-th particle, respectively. A general position-and-velocity dependent force $F_i(\textit{\textbf{x}}, \textit{\textbf{v}})$ is applied to the $i$-th particle, where $\textit{\textbf{x}}=(x_1,\cdots,x_N)$ and $\textit{\textbf{v}}=(v_1,\cdots,v_N)$. The dynamics of the $i$-th particle is described by the following equation:
\begin{align}
\dot{x}_i = v_i,~~m_i \dot{v}_i = F_i(\textit{\textbf{x}}, \textit{\textbf{v}}) - \gamma_i v_i +\xi_i, \label{eq:underdamped_Langevin}
\end{align}
where $m_i$, $\gamma_i$, and $\xi_i$ are the mass, the damping coefficient, and the Gaussian white noise, satisfying $\langle \xi_i (t) \xi_j (t^\prime) \rangle = 2 k_\textrm{B} \gamma_i T_i \delta_{ij} \delta(t-t^\prime) $, respectively. For brevity, we set $k_\textrm{B} =1$ for the following discussion.
If we define $P(\textit{\textbf{x}}, \textit{\textbf{v}},t)$ as the probability distribution function, this dynamics can be also described by the following Fokker-Planck equation:
\begin{align}
\partial_t P(\textit{\textbf{x}}, \textit{\textbf{v}},t) = - \sum_{i=1}^N (\partial_{x_i} J_{x_i} + \partial_{v_i} J_{v_i}), \label{eq:underdamped_FP}
\end{align}
where the probabilistic currents $J_{x_i} = v_i P(\textit{\textbf{x}}, \textit{\textbf{v}},t)$ and $J_{v_i} = m_i^{-1} [-\gamma_i v_i +F_i(\textit{\textbf{x}}, \textit{\textbf{v}}) - T_i \gamma_i m_i^{-1} \partial_{v_i} ] P(\textit{\textbf{x}}, \textit{\textbf{v}},t) $.

Now we consider a single trajectory in the $(\textit{\textbf{x}}, \textit{\textbf{v}})$ phase space of this dynamics from time $t=0$ to $t=\mathcal{T}$, which is denoted by $\Gamma \equiv [\textit{\textbf{x}}(t), \textit{\textbf{v}}(t)]_{t=0}^{t=\mathcal{T}}$. Note that $(\textit{\textbf{x}}_0, \textit{\textbf{v}}_0)$ is  the starting point of this trajectory. The probability density observing the trajectory  $\Gamma$ is denoted by $\mathcal{P}[\Gamma]$.
To calculate the EP (or irreversibility), we need to define the time-reverse dynamics~\cite{Udo_review}. The time-reverse position and velocity variables, $\tilde{\textit{\textbf{x}}}$ and $\tilde{\textit{\textbf{v}}}$, should satisfy
\begin{align}
\dot{\tilde{x}}_i = \tilde{v}_i,~~m_i \dot{\tilde{v}}_i = F_i^\dagger(\tilde{\textit{\textbf{x}}}, \tilde{\textit{\textbf{v}}}) - \gamma_i \tilde{v}_i +\xi_i, \label{eq:Langevin_eq_reversal}
\end{align}
where the $\dagger$ operation reverses signs of all odd parameters in the time-reversal process. $\mathcal{P}^\dagger [\tilde{\Gamma}]$ is the probability density observing
the time-reverse trajectory $\tilde{\Gamma} \equiv [\textit{\textbf{x}}(\mathcal{T}-t), -\textit{\textbf{v}}(\mathcal{T}-t)]_{t=0}^{t=\mathcal{T}}$ in the $\dagger$ dynamics of Eq.~\eqref{eq:Langevin_eq_reversal}. Note that there is no unique way to choose odd parameters for the time-reversal process. For example, one may regard a magnetic field as an odd parameter, so change the sign of a magnetic field in the time-reverse dynamics~\cite{Andreas2,ChunNoh}. On the other hand, one may keep the sign of the magnetic field for the
irreversibility~\cite{spinney,spinney1,lkp,Kwon,Yeo}, where the $\dagger$ dynamics is identical to the original time-forward dynamics. Nonetheless, we will show later that the generalized TUR does not depend on the choice of odd parameters.

With a certain choice of odd parameters, we divide the force into the reversible and irreversible one as
\begin{align}
F_i(\textit{\textbf{x}}, \textit{\textbf{v}}) = F_i^\textrm{rev}(\textit{\textbf{x}}, \textit{\textbf{v}}) + F_i^\textrm{ir}(\textit{\textbf{x}}, \textit{\textbf{v}}),
\end{align}
with
$F_i^\textrm{rev}(\textit{\textbf{x}}, \textit{\textbf{v}}) = {F_i^\textrm{rev}}^\dagger (\textit{\textbf{x}}, -\textit{\textbf{v}})$, and $F_i^\textrm{ir}(\textit{\textbf{x}}, \textit{\textbf{v}}) = - {F_i^\textrm{ir}}^\dagger (\textit{\textbf{x}}, -\textit{\textbf{v}})$.
Then, we get the irreversible part of the velocity component of the probability current as
\begin{align}
J_{v_i}^\textrm{ir} = \frac{1}{m_i} \left[ F_i^\textrm{ir} (\textit{\textbf{x}}, \textit{\textbf{v}} ) - \gamma_i v_i -\frac{T_i \gamma_i}{m_i} \partial_{v_i} \right] P(\textit{\textbf{x}}, \textit{\textbf{v}},t) . \label{eq:irreversible_current}
\end{align}
The total EP is determined by the ratio between the two trajectory probabilities $\mathcal{P}[\Gamma]$ and $\mathcal{P}^\dagger(\tilde{\Gamma})$, that is, $\Delta S_\textrm{tot} =  \ln [\mathcal{P}[\Gamma] /\mathcal{P}^\dagger[\tilde{\Gamma}] ] $ ~\cite{Udo_review}. In addition, the average EP rate can be written as~\cite{spinney,Kwon}:
\begin{align}
\sigma \equiv \langle \dot{S}_\textrm{tot} \rangle = \sum_{i=1}^N \frac{m_i^2}{T_i \gamma_i} \int d\textit{\textbf{x}} \int  d \textit{\textbf{v}} \frac{(J_{v_i}^\textrm{ir})^2}{P(\textit{\textbf{x}}, \textit{\textbf{v}},t)}\label{EP_rate}
\end{align}

The main goal of this study is finding a generalized TUR for a general current $\Theta$ which has the following form:
\begin{align}
\Theta[\Gamma] = \int_0^\mathcal{T} dt {\bf \Lambda} (\textit{\textbf{x}} (t), \textit{\textbf{v}}(t) ) \circ \textit{\textbf{v}}(t),
\end{align}
where ${\bf \Lambda}$ is an arbitrary $N$-dimensional vector and $\circ$ denotes the Stratonovich multiplication.
To go further, we take the virtual perturbation approach by Vu and Hasegawa~\cite{Dechant1,Vu1,Vu2}. First, consider an auxiliary dynamics
\begin{align}
\dot{x}_i = v_i,~~m_i \dot{v}_i = H_{\theta,i} (\textit{\textbf{x}}, \textit{\textbf{v}})+\xi_i, \label{eq:auxiliary_eq}
\end{align}
where $H_{\theta,i}$ is an auxiliary force with $H_{\theta=0,i} = F_i -\gamma_i v_i$. Thus, the auxiliary dynamics~\eqref{eq:auxiliary_eq} becomes the original dynamics~\eqref{eq:underdamped_Langevin} at $\theta=0$. The detailed form of $H_{\theta,i}$ will be given later. The trajectory probability density  in the auxiliary dynamics is denoted by $\mathcal{P}_\theta [\Gamma]$. In the Onsager-Machlup formulation~\cite{Onsager}, $\mathcal{P}_\theta [\Gamma]$ in the Ito scheme is given by
\begin{align}
\mathcal{P}_\theta[\Gamma] = \mathcal{N} P_\theta (\textit{\textbf{x}}_0, \textit{\textbf{v}}_0)  \prod_{i=1}^N \exp[-\mathcal{A}_i[\Gamma]],
\end{align}
where the action $\mathcal{A}_i [\Gamma] = \int_0^\mathcal{T} dt \frac{1}{4 T_i \gamma_i} (m_i \dot{v}_i - H_{\theta,i} (\textit{\textbf{x}}, \textit{\textbf{v}}) )^2$ and $\mathcal{N}$ is the normalization constant which is independent of $\theta$.

The trajectory-ensemble average of $\Theta[\Gamma]$ in the auxiliary dynamics is  $\langle \Theta \rangle_\theta
= \int \mathcal{D}\Gamma \Theta[\Gamma] \mathcal{P}_\theta [\Gamma]$.
With $\textrm{Var}_\theta [\Theta] \equiv \langle \Theta^2\rangle_\theta -\langle \Theta \rangle_\theta^2$, the Cram\'er-Rao inequality yields~\cite{Vu2}
\begin{align}
\frac{\textrm{Var}_\theta [\Theta]}{ (\partial_\theta \langle \Theta\rangle_\theta )^2 } \geq \frac{1}{\mathcal{I}(\theta)}, \label{eq:CramerRao}
\end{align}
where $\mathcal{I}$ is the Fisher information given by
\begin{align}
\mathcal{I}(\theta) = - \langle \partial_\theta^2 \ln P_\theta (\textit{\textbf{x}}_0, \textit{\textbf{v}}_0)  \rangle_\theta + \frac{1}{2} \left\langle \sum_{i=1}^N \int_0^\mathcal{T} dt \frac{(\partial_\theta H_{\theta,i})^2}{T_i \gamma_i} \right\rangle_\theta. \label{eq:FisherInformation}
\end{align}
We slightly modify the perturbation technique presented in Ref.~\cite{Vu1} in order to include a velocity-dependent force by considering the auxiliary force as
\begin{align}
H_{\theta,i} (\textit{\textbf{x}}, \textit{\textbf{v}}) =& -(1+\theta) \gamma_i v_i + (1+\theta)^2 F_i \left( \textit{\textbf{x}}, \frac{\textit{\textbf{v}} }{1+\theta} \right) \nonumber \\
&+ \frac{T_i \gamma_i}{m_i} \left(1-(1+\theta)^3 \right) \frac{ \partial_{v_i} P^\textrm{ss} \left( \textit{\textbf{x}}, \frac{\textit{\textbf{v}} }{1+\theta} \right)  }{P^\textrm{ss} \left(\textit{\textbf{x}}, \frac{\textit{\textbf{v}} }{1+\theta} \right) }, \label{eq:auxiliary_force}
\end{align}
where $P^\textrm{ss} \left(\textit{\textbf{x}}, \textit{\textbf{v}}  \right)$ is the steady-state solution of the original dynamics~\eqref{eq:underdamped_FP}.
By using Eq.~\eqref{eq:auxiliary_force}, it can be easily shown that the steady-state probability distribution function of the Fokker-Planck equation of the auxiliary dynamics is given by
\begin{align}
P_\theta^\textrm{ss} (\textit{\textbf{x}}, \textit{\textbf{v}}  ) = \frac{ P^\textrm{ss} (\textit{\textbf{x}},  \frac{ \textit{\textbf{v}} }{1+\theta}  ) }{(1+\theta)^N}.
\end{align}
For a general ${\bf \Lambda} (\textit{\textbf{x}} (t), \textit{\textbf{v}}(t) )$, $\partial_\theta \langle \Theta\rangle_\theta$ at $\theta=0$ in the steady state becomes
\begin{align}
\partial_\theta \langle \Theta\rangle_{\theta, \textrm{s}} |_{\theta = 0} &= \partial_\theta \left. \left(  \mathcal{T} \int d\textit{\textbf{x}} \int  d \textit{\textbf{v}}   {\bf \Lambda} (\textit{\textbf{x}} , \textit{\textbf{v}} ) \cdot \textit{\textbf{v}}  P_\theta^\textrm{ss}  (\textit{\textbf{x}} , \textit{\textbf{v}} ) \right) \right|_{\theta =0} \nonumber \\
&= \mathcal{T} \int d\textit{\textbf{x}} \int  d \textit{\textbf{v}}    P^\textrm{ss}  (\textit{\textbf{x}} , \textit{\textbf{v}} ) \textit{\textbf{v}} \cdot (1+ \textit{\textbf{v}} \cdot \nabla_\textit{\textbf{v}}) {\bf \Lambda} (\textit{\textbf{x}} , \textit{\textbf{v}} ) \nonumber \\
&= \langle \Theta \rangle_\textrm{s} + \langle \Theta^\prime \rangle_\textrm{s} \label{eq:Theta}
\end{align}
where $ \langle \Theta \rangle_\textrm{s} = \mathcal{T} \int  d\textit{\textbf{x}}  \int  d \textit{\textbf{v}}   {\bf \Lambda} (\textit{\textbf{x}} , \textit{\textbf{v}} )  \cdot  \textit{\textbf{v}}  P^\textrm{ss}  (\textit{\textbf{x}} , \textit{\textbf{v}} ) $ and $ \langle \Theta^\prime \rangle_\textrm{s} = \mathcal{T} \int  d\textit{\textbf{x}}  \int  d \textit{\textbf{v}}  \left[ (\textit{\textbf{v}} \cdot \nabla_\textit{\textbf{v}}) {\bf \Lambda} (\textit{\textbf{x}} , \textit{\textbf{v}} ) \right] \cdot  \textit{\textbf{v}}  P^\textrm{ss}  (\textit{\textbf{x}} , \textit{\textbf{v}} ) $.

Now, we calculate the Fisher information $\mathcal{I}(\theta)$ at $\theta=0$.
The second bulk term in Eq.~\eqref{eq:FisherInformation} can be split into the term proportional to the EP and the term proportional to the dynamic activity, while
the first term yields the boundary term independent of the time duration $\mathcal{T}$.
After some algebra in the steady state, we arrive at
\begin{align}
\mathcal{I} (0) = \frac{1}{2} \left[  \mathcal{T}(9 \sigma_\textrm{s} + 4 \Upsilon_\textrm{uv}) +\Omega\right], \label{eq:FisherAtZero}
\end{align}
where $\sigma_\textrm{s}$  is the steady-state EP rate  given by Eq.~\eqref{EP_rate},
$\Upsilon_\textrm{uv}$  is the {\em generalized} dynamic activity, and $\Omega$ is the boundary term,
which are expressed by
\begin{align}
\Upsilon_\textrm{uv} =& \sum_{i=1}^N \left( -\frac{2\gamma_i}{T_i}\langle v_i^2\rangle_\textrm{s}  -\frac{1}{2T_i}\langle v_i \mathcal{F}_i \rangle_\textrm{s} + \frac{1}{4T_i \gamma_i }\langle \mathcal{F}_i^2\rangle_\textrm{s} \right. \nonumber \\
& \left. +\frac{3}{2T_i \gamma_i }\langle \mathcal{F}_i F_i^\textrm{ir} \rangle_\textrm{s} +\frac{3}{T_i }\langle v_i F_i^\textrm{ir} \rangle_\textrm{s} +\frac{3}{2 m_i  }\langle \partial_{v_i} \mathcal{F}_i  \rangle_\textrm{s} +\frac{3 \gamma_i}{m_i} \right), \nonumber\\
\Omega =& 2 \left\langle  \left( \frac{\sum_{i=1}^N v_i  \partial_{v_i} P^\textrm{ss} (\textit{\textbf{x}}, \textit{\textbf{v}}) }{P^\textrm{ss} (\textit{\textbf{x}}, \textit{\textbf{v}})} \right)^2 \right\rangle_\textrm{s}  -2 N^2, \label{eq:Upsilon_Omega}
\end{align}
with
\begin{align}
\mathcal{F}_i = 2F_i^\textrm{rev} - F_i^\textrm{ir} - v_i \partial_{v_i} F_i.
\end{align}

From Eqs.~\eqref{eq:CramerRao}, \eqref{eq:Theta}, and \eqref{eq:FisherAtZero},
the generalized TUR is written as
\begin{align}
\mathcal{Q}_\textrm{uv} (\Theta) \equiv \frac{\textrm{Var}[\Theta]}{\langle \Theta \rangle_\textrm{s}^2} \Sigma_\textrm{uv} \geq 2,~~~\Sigma_\textrm{uv} \equiv \frac{\mathcal{T}(9 \sigma_\textrm{s} + 4 \Upsilon_\textrm{uv}) +\Omega}{ (1+\langle \Theta^\prime \rangle_\textrm{s}/\langle \Theta \rangle_\textrm{s} )^2} \label{eq:main_ineq}
\end{align}
where the subscript `uv' represents the `underdamped dynamics with a velocity dependent force'.
The main difference of Eq.~\eqref{eq:main_ineq} from Eq.~\eqref{eq:underdampedQ}
is the form of the dynamical activity. This difference  goes away
without a velocity-dependent force by setting $\partial_{v_i} F_i=0$ and $ F_i^\textrm{ir}=0$.
It should be emphasized that  $\Sigma_\textrm{uv} $ is independent of the choice of odd
parameters, even though the EP rate $\sigma_\textrm{s}$ and the dynamic activity $\Upsilon$ depends on the choice, respectively. This implies that the original TUR depends on this choice.
We also note that, when  ${\bf \Lambda}$ has only position-dependent terms, $\langle \Theta^\prime \rangle_\textrm{s} =0$. If $ {\bf \Lambda}_i (\textit{\textbf{x}} , \textit{\textbf{v}} ) = \lambda_i(\textit{\textbf{x}}) v_i^d $, $\langle \Theta^\prime \rangle_\textrm{s} / \langle \Theta \rangle_\textrm{s} = d$. These cases had been previously discussed in Ref.~\cite{Vu1}.

\section {Examples}
In this section, we test our generalized TUR~\eqref{eq:main_ineq} for three systems affected by velocity-dependent forces: (i) a molecular refrigerator, (ii) a Brownian particle in a magnetic field with a rotational force, and (iii) an active matter system.

\subsection{Molecular refrigerator}

Here, we consider a one-dimensional Brownian particle immersed in a reservoir with temperature $T$. Its position and velocity are denoted by $x$ and $v$, respectively. An external linear dissipative force $-\alpha v$ is applied to the particle. Then, the equation of motion becomes
\begin{align}
\dot{x} = v, ~~ m\dot{v} = - \alpha v - \gamma v + \xi, \label{eq:MolRef}
\end{align}
where $\gamma$ and $\xi$ are the damping coefficient and the Gaussian white noise satisfying $\langle \xi(t) \xi(t^\prime) \rangle = 2 \gamma T \delta(t-t^\prime)$, respectively. Note that $\alpha+\gamma >0$ for the stability of the dynamics. Equation~\eqref{eq:MolRef} describes the motion of the simplest molecular refrigerator~\cite{Liang, Cohadon, Pinard, Mertz, Jourdan, Kim}. The external dissipative force reduces the thermal fluctuation of the particle when $\alpha >0$. Thus, its motion is effectively the same as that of a particle in a cooler environment, which mimics a refrigerator at the particle level. This mechanism is often used for reducing thermal fluctuations of a mesoscopic system such as a suspended mirror of interferometric detectors~\cite{Cohadon, Pinard} and an atomic-force-microscope (AFM) cantilever~\cite{Mertz, Jourdan}.

The steady state of Eq.~\eqref{eq:MolRef} is given by the Boltzmann distribution, that is,
\begin{align}
P^\textrm{ss}(v) = \sqrt{\frac{m}{ 2\pi T^\prime }} \exp\left( -\frac{m}{2T^\prime} v^2 \right), \label{def:ssMolRef}
\end{align}
where $T^\prime = \gamma T /(\gamma + \alpha)$ is the effective temperature.

In this problem, we choose the reversible and irreversible forces as follows:
\begin{align}
F^\textrm{rev}  =0,~~ F^\textrm{ir} = - \alpha v, ~~\mathcal{F} = 2\alpha v.
\end{align}
Then, it is straightforward to show that $\Upsilon_\textrm{uv} =(\alpha+\gamma)/m$ from the fact $\langle v^2\rangle_\textrm{s} = T^\prime/m$. In addition, the irreversible current becomes zero from Eq.~\eqref{eq:irreversible_current}, thus $\sigma_\textrm{s} =0$. Finally, $\Omega = 4$.
Therefore, $\Sigma_\textrm{uv}$ of the molecular refrigerator becomes
\begin{align}
\Sigma_\textrm{uv}^\textrm{mr} = \left( 4\mathcal{T} \frac{\alpha +\gamma}{m} +4 \right)/(1+\langle \Theta^\prime \rangle_\textrm{s}/\langle \Theta \rangle_\textrm{s})^2. \label{eq:Sigma_MolRef}
\end{align}
Note that we can choose another reversible and irreversible force as $F^\textrm{rev} =  -\alpha v$ and $F^\textrm{ir}=0$, respectively~\cite{Kim}. In this case, the EP rate changes as $\sigma_\textrm{s} =\alpha^2/[m(\alpha+\gamma)]$ (called as entropy pumping~\cite{Kim}), while $\Sigma_\textrm{uv}^\textrm{mr}$ remains unchanged.

\begin{figure}
\centering
\includegraphics[width=\linewidth]{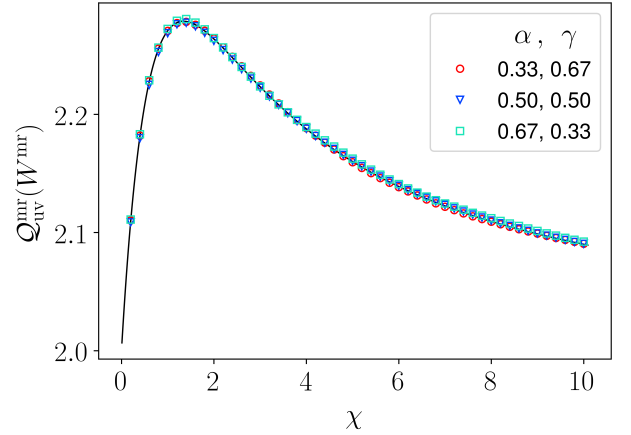}
\caption{The TUR factor $\mathcal{Q}$ for the molecular refrigerator as a function of the rescaled observation time $\chi$.
Red circles, blue down-triangles, and
cyan squares are the numerical results
for $(\alpha, \gamma) = (0.33, 0.67)$,
$(0.5, 0.5)$, and $(0.67, 0.33)$, respectively. For this calculation,
$m=1$ and $T=1$ are used. The black solid curve denotes the analytic result~\eqref{eq:Qmr}.
} \label{fig:mr}
\end{figure}

To test the validity of Eq.~\eqref{eq:main_ineq}, we choose the work current for $\Theta$
with $\Lambda (v)=-\alpha v$. It is easy to calculate the average work  in the steady state as
\begin{align}
\langle W^\textrm{mr}  \rangle_\textrm{s} = \mathcal{T} \langle -\alpha v^2\rangle_\textrm{s} = -\frac{\alpha \gamma T }{m(\alpha+\gamma)} \mathcal{T}. \label{eq:MolRef_W}
\end{align}
Its variance can be also calculated explicitly (see the detailed calculation in Appendix A):
\begin{align}
\textrm{Var} [W^\textrm{mr}] = \frac{2 \alpha^2 \gamma^2 T^2}{m(\alpha + \gamma)^3} \left( \mathcal{T} - \frac{m[1-\exp(-2(\alpha+\gamma)\mathcal{T}/m) ]}{2(\alpha +\gamma)}  \right).
\label{eq:MolRef_Wvar}
\end{align}
Combining Eqs.~\eqref{eq:MolRef_W}, \eqref{eq:MolRef_Wvar}, and \eqref{eq:Sigma_MolRef} with the fact $\langle {W^\textrm{mr}}^\prime \rangle_\textrm{s}/\langle W^\textrm{mr} \rangle_\textrm{s}=1$, we find the TUR factor for the work  as
\begin{align}
\mathcal{Q}_\textrm{uv}^\textrm{mr} (W^\textrm{mr}) \equiv \frac{\textrm{Var}[W^\textrm{mr}]}{\langle W^\textrm{mr} \rangle^2 } \Sigma_\textrm{uv}^\textrm{mr} = 2\left(1+\frac{1}{\chi} \right)\left(1-\frac{1-e^{-2 \chi}}{2\chi} \right), \label{eq:Qmr}
\end{align}
where $\chi =\mathcal{T}/\tau_\textrm{relax}$  is the observation time in the unit of the relaxation time $\tau_\textrm{relax}\equiv m/(\alpha+\gamma)$.
This factor turns out to be always larger than 2, as expected (see Fig.~\ref{fig:mr}).
It is noteworthy to mention that the lowest bound ($\mathcal{Q}_\textrm{uv}\rightarrow 2$) is reachable in both the $\chi \rightarrow 0$ (short-time) and $\chi \rightarrow \infty$ (long-time) limit. Thus, our generalized TUR provides a {\em tight} bound for work fluctuations in this case.

To confirm our analysis, we performed numerical calculations by solving Eq.~\eqref{eq:MolRef} using a generalized velocity-Verlet algorithm~\cite{Ciccotti} which is correct up to the second order of simulation time step $\Delta t=0.01$.
We obtained $\mathcal{Q}_\textrm{uv}^\textrm{mr} (W^\textrm{mr})$ by averaging over $10^7$ sample paths starting from random initial states sampled from the steady-state distribution~\eqref{def:ssMolRef}. Numerical results are shown in Figure~\ref{fig:mr},
which  are perfectly matched to the analytic curve~\eqref{eq:Qmr}.


We check whether the original TUR \eqref{eq:originalQ} is valid or not in this model.
As mentioned previously, the EP rate depends on the choice of the odd parameters.
With the choice of $F^\textrm{rev}  =0$ and $ F^\textrm{ir} = - \alpha v$, we get $\sigma_\textrm{s}=0$. Therefore, the original TUR factor $\mathcal{Q}_\textrm{ori}^\textrm{mr} (W^\textrm{mr}) = 0$, which clearly violates
the original TUR. 
With the choice of $F^\textrm{rev}  =- \alpha v$ and $ F^\textrm{ir} = 0$, we get $\sigma_\textrm{s} =\alpha^2/[m(\alpha+\gamma)]$, leading to
\begin{align}
\mathcal{Q}_\textrm{ori}^\textrm{mr} (W^\textrm{mr}) = \frac{2\alpha^2}{(\alpha+\gamma)^2  }\left( 1 - \frac{1-e^{-2\chi}}{2 \chi} \right) . \label{eq:Qri_MolRef}
\end{align}
When $\alpha>0$, $\mathcal{Q}_\textrm{ori}^\textrm{mr} (W^\textrm{mr})$
becomes smaller than $2$. Again, the original TUR is broken.

Furthermore, the modified TUR found by Vu and Hasegawa~\eqref{eq:underdampedQ}
does not hold either. Their dynamical activity is given by~\cite{Vu1}
\begin{align}
\Upsilon = \frac{1}{T \gamma} \langle F^2 \rangle_\textrm{s} - 3 \frac{\gamma}{T} \langle v^2 \rangle_\textrm{s} + 4\frac{\gamma}{m} = \frac{\gamma^2 +\alpha^2 + 4\gamma \alpha }{m(\gamma + \alpha)},
\end{align}
where $F=-\alpha v$.
The EP rate is choice-dependent as discussed above.
It turns out that any choice yields a negative value for $\Sigma_\textrm{u}$ in Eq.~\eqref{eq:underdampedQ} for some parameter range
of $\alpha$, which clearly demonstrates the violation of Eq.~\eqref{eq:underdampedQ}.

\subsection{Brownian particle in a magnetic field with a rotational force}

Recently, Chun et al.~\cite{Chun} studied the validity of the original TUR for a Brownian particle in a magnetic field. They found that the original TUR on the work current can be broken due to a magnetic field. To find a correct uncertainty relation, we should take account of the effect of a velocity-dependent force, since the magnetic field induces a Lorentz force.

To check the generalized TUR, we consider the same system studied by Chun et al.~\cite{Chun}. Suppose that a charged Brownian particle is trapped in a harmonic potential with stiffness $k$ and immersed in a reservoir with temperature $T$. The particle moves in a two-dimensional space, and its position and velocity are denoted by $\textit{\textbf{x}} = (x_1, x_2)$ and $\textit{\textbf{v}} = (v_1, v_2)$, respectively. The Lorentz force $(B v_2, -B v_1)$ and a nonconservative rotational force $(\kappa x_2, -\kappa x_1)$ are applied to the particle. Then, the equation of motion can be written as
\begin{align}
\dot{x}_1 &=v_1,~~m\dot{v}_1 = B v_2 -kx_1 + \kappa x_2 - \gamma v_1 +\xi_1, \nonumber \\
\dot{x}_2 &=v_2,~~m\dot{v}_2 = -B v_1 -kx_2 - \kappa x_1 - \gamma v_2 +\xi_2. \label{eq:EoM_Bfield}
\end{align}
In this dynamics, the total forces are $F_1 = Bv_2 - kx_1 + \kappa x_2$ and $F_2 = -B v_1 -kx_2 -\kappa x_1$. We choose the reversible part of $F_i$ as $F_1^\textrm{rev} = F_1$ and $F_2^\textrm{rev} = F_2$, with the irreversible parts $F_1^\textrm{ir} = 0$ and $F_2^\textrm{ir} = 0$. Then, we get $\mathcal{F}_1 = 2F_1$ and $\mathcal{F}_2 =2F_2$. With this choice, the irreversible current~\eqref{eq:irreversible_current} simply becomes the heat current~\cite{ChunNoh,Kwon} and the EP rate is given by
\begin{align}
\sigma_\textrm{s} = -\frac{ \langle \dot{Q}_1 \rangle_\textrm{s} + \langle \dot{Q}_2 \rangle_\textrm{s} }{T} = -\frac{2\gamma}{m} +\frac{\gamma}{T} \left( \langle v_1^2 \rangle_\textrm{s} +\langle v_2^2 \rangle_\textrm{s} \right), \label{eq:EPrate_Bfield}
\end{align}
where $\dot{Q}_i  =  (-\gamma v_i +\xi_i)\circ v_i$ is the heat current in the $i$-th direction. The second equality of Eq.~\eqref{eq:EPrate_Bfield} comes from the fact $\langle \dot{Q}_i \rangle_\textrm{s} = \gamma (T/m - \langle v^2 \rangle_\textrm{s} )$~\cite{Kwon,Park} which can be straightforwardly shown by using the Stratonovich multiplication.

For evaluating the boundary term $\Omega$, we need to know the distribution function $P^\textrm{ss} (\textit{\textbf{x}}, \textit{\textbf{v}})$. For this purpose, we rewrite Eq.~\eqref{eq:EoM_Bfield} in the form of a multivariate Ornstein-Uhlenbeck process as
\begin{align}
\dot{\textit{\textbf{z}}} = -\textsf{A} \textit{\textbf{z}}  + {\bm  \eta},
\end{align}
where $\textit{\textbf{z}} = (x_1,x_2,v_1,v_2)^\textsf{T}$, ${\bm \eta} = (0,0,\xi_1,\xi_2)^\textsf{T}$, and
\begin{align}
\textsf{A} = \frac{1}{m}\begin{pmatrix}
0 & 0 & -m & 0 \\
0 & 0 & 0 & -m \\
k & -\kappa & \gamma & -B \\
\kappa & k & B & \gamma
 \end{pmatrix}.
\end{align}
As we are interested in the steady state, we need a condition for guaranteeing a stable solution of this linear system which is given by~\cite{Chun,LeeS}
\begin{align}
K\equiv k+\frac{\kappa B}{\gamma} -\frac{\kappa^2 m}{\gamma^2} >0. \label{eq:stability_Bfield}
\end{align}
We define the covariance matrix $\textsf{C} \equiv \langle \textit{\textbf{z}} \textit{\textbf{z}}^\textsf{T} \rangle_\textrm{s} $ which is the solution of the matrix equation $\textsf{A} \textsf{C} + \textsf{C} \textsf{A}^\textsf{T} = 2 \textsf{D}$, where $\textsf{D} =  (\gamma T/m^2) \textrm{diag}\{ 0,0,1,1 \} $. The solution is~\cite{Gadiner,Chun,LeeS}
\begin{align}
\textsf{C} = \frac{T}{\gamma K}\begin{pmatrix}
\gamma & 0 & 0 & -\kappa \\
0 & \gamma & \kappa & 0 \\
0 & \kappa & (\gamma k +\kappa B)/m & 0 \\
-\kappa & 0 & 0 & (\gamma k +\kappa B)/m
 \end{pmatrix}.
\end{align}
Then, the distribution function can be written as~\cite{Chun}
\begin{align}
P^\textrm{ss} (\textit{\textbf{z}}) = \frac{1}{\sqrt{ \textrm{det}2\pi \textsf{C}  } } \exp\left[ -\frac{1}{2} \textit{\textbf{z}}^\textsf{T} \cdot \textsf{C}^{-1}  \textit{\textbf{z}} \right]. \label{eq:Pss_Bfield}
\end{align}
By inserting Eq.~\eqref{eq:Pss_Bfield} to Eq.~\eqref{eq:Upsilon_Omega}, we can explicitly calculate $\Omega$. The result is
\begin{align}
\Omega = 8 + \frac{4m \kappa^2}{\gamma^2 K}. \label{eq:Omega_Bfield}
\end{align}
Finally, using the fact $\langle x_1^2\rangle_\textrm{s} =\textsf{C}_{11}$, $\langle x_2^2\rangle_\textrm{s} =\textsf{C}_{22}$, $\langle v_1^2\rangle_\textrm{s} =\textsf{C}_{33}$, $\langle v_2^2\rangle_\textrm{s} =\textsf{C}_{44}$, $\langle x_1 v_2 \rangle_\textrm{s} =\textsf{C}_{14}$, and $\langle x_2 v_1 \rangle_\textrm{s} =\textsf{C}_{23}$, the dynamical activity can be written as
\begin{align}
\Upsilon_\textrm{uv}^\textrm{mag} =& \frac{B^2-2\gamma^2}{\gamma T} (\textsf{C}_{33}+\textsf{C}_{44}) +\frac{k^2 +\kappa^2}{\gamma T} (\textsf{C}_{11}+ \textsf{C}_{22}) \nonumber \\
&+\frac{\kappa \gamma-2Bk}{\gamma T} (\textsf{C}_{14}-\textsf{C}_{23}) +\frac{6\gamma}{m}. \label{eq:Upsilon_Bfield}
\end{align}
Combining Eqs.~\eqref{eq:EPrate_Bfield}, \eqref{eq:Omega_Bfield}, and \eqref{eq:Upsilon_Bfield}, we have consequently
\begin{align}
\Sigma_\textrm{uv}^\textrm{mag} =
\frac{\mathcal{S} \mathcal{T} +8+\frac{4m \kappa^2}{\gamma^2 K}}{(1+\langle \Theta^\prime \rangle_\textrm{s}/\langle \Theta \rangle_\textrm{s})^2} , \label{eq:SigmaBfield}
\end{align}
where the bulk term $\mathcal{S}=9 \sigma_\textrm{s}+4  \Upsilon_\textrm{uv}^\textrm{mag}$ is given as
\begin{equation}
\mathcal{S} =  \frac{8\gamma (B^2 +\gamma^2)K +2m\gamma [4(K +\kappa^2 m\gamma^{-2})^2+\kappa^2]}{m\gamma^2 K}.
\end{equation}

We test our generalized TUR by taking the accumulated work
$W^\textrm{mag}$ done by the nonconservative rotational force,
that is, with ${\bf \Lambda}(\textit{\textbf{x}} )=(\kappa x_2,-\kappa x_1)$ as
\begin{align}
W^\textrm{mag}(\mathcal{T}) =\int_0^\mathcal{T}  \kappa \left[ x_2(t) v_1(t) - x_1(t) v_2(t) \right] dt.
\end{align}
As $ {\bf \Lambda}$ has only position-dependent terms, $\langle {W^\textrm{mag}}^\prime \rangle_\textrm{s}/\langle W^\textrm{mag} \rangle_\textrm{s}=0$.
The long-time behavior of
the relative fluctuation of the accumulated work was already calculated in Ref.~\cite{Touchette,Chun}, which is
\begin{align}
\frac{2 D_W^\textrm{mag}}{ \langle \dot{W}^\textrm{mag} \rangle_\textrm{s}^2} = \frac{ \left[ 1+ \kappa_0^2 (1+3 m_0) + \kappa_0^3 m_0 B_0 \right] \gamma  }{(1+\kappa_0 B_0 - \kappa_0^2 m_0) \kappa_0^2 k}, \label{eq:WorkFlucBfield}
\end{align}
where the diffusion coefficient $D_W^\textrm{mag}$ is defined as
\begin{align}\label{def:Dw_mag}
    D_W^\textrm{mag}
    \equiv \lim_{\mathcal{T} \rightarrow \infty}
    \frac{\textrm{Var} [W^\textrm{mag}]}{2 \mathcal{T}}
\end{align}
and $\kappa_0 = \kappa/k$, $B_0 = B/\gamma$, and $m_0 = km /\gamma^2 $ are dimensionless parameters. By multiplying Eqs.~\eqref{eq:SigmaBfield} and \eqref{eq:WorkFlucBfield},  we finally get
\begin{align}\label{def:Q_uv_inf_mag}
\mathcal{Q}_{\textrm{uv},\infty}^\textrm{mag} (W^\textrm{mag}) \equiv \frac{2 D_W^\textrm{mag}}{ \langle \dot{W}^\textrm{mag} \rangle_\textrm{s}^2} \mathcal{S},
\end{align}
where the subscript `$\infty$' denotes that $\mathcal{Q}_\textrm{uv}^\textrm{mag}$ is evaluated in the $\mathcal{T} \rightarrow \infty$ limit.

\begin{figure}
\centering
\includegraphics[width=\linewidth]{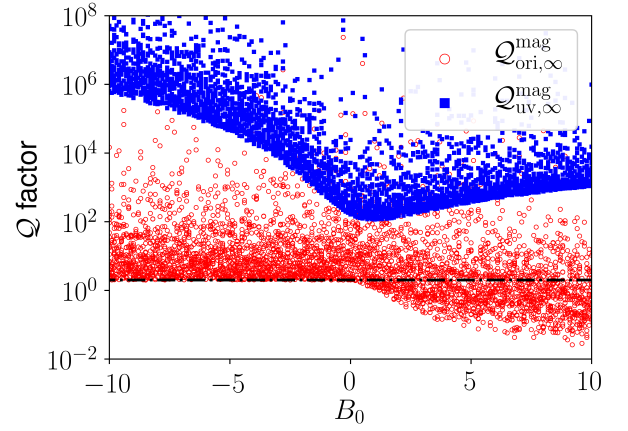}
\caption{The TUR factor $\mathcal{Q}$  as a function of the rescaled magnetic field $B_0$ in the infinite time limit for a charged Brownian particle in a magnetic field. Open red circles represent the results calculated from the original TUR~\eqref{def:Q_ori_inf_mag}, whereas filled blue squares represent those from our generalized TUR~\eqref{def:Q_uv_inf_mag}. Under the stability condition~\eqref{eq:stability_Bfield},
we choose the parameters $m_0 \in [0, 1]$ and $\kappa_0 \in [0, 10]$ for a given $B_0$ to plot data points. The black dashed-dotted line indicates the TUR bound, that is, $\mathcal{Q} =2$.
} \label{fig:Bfield}
\end{figure}

To confirm our bound, we numerically calculate $\mathcal{Q}_{\textrm{uv},\infty}^\textrm{mag} (W^\textrm{mag})$ for many parameter sets of $(\kappa_0, B_0, m_0)$  and plot them in Fig.~\ref{fig:Bfield}. As shown in the figure, $\mathcal{Q}_{\textrm{uv},\infty}^\textrm{mag} (W^\textrm{mag})$ satisfies our bound~\eqref{eq:main_ineq}. However, it seems that the lowest bound is not reachable at all in this case; the minimum value of $\mathcal{Q}_{\textrm{uv},\infty}^\textrm{mag} (W^\textrm{mag})$ is about $\sim 100$, which indicates that $\mathcal{Q}_{\textrm{uv},\infty}^\textrm{mag} (W^\textrm{mag})$ provides a very {\em loose} (so not much useful) bound in this system. For comparison, we also plot the values of the original TUR
factor $\mathcal{Q}_{\textrm{ori},\infty}^\textrm{mag} (W^\textrm{mag})$ for many parameter sets, which is given by
\begin{align}\label{def:Q_ori_inf_mag}
\mathcal{Q}_{\textrm{ori},\infty}^\textrm{mag} (W^\textrm{mag}) \equiv \frac{2 D_W^\textrm{mag}}{ \langle \dot{W}^\textrm{mag} \rangle_\textrm{s}^2} \sigma_\textrm{s},
\end{align}
where we calculate the EP rate $\sigma_\textrm{s}$  as in Eq.~\eqref{eq:EPrate_Bfield}.
As already reported in Ref.~\cite{Chun}, the origianl TUR is satisfied as $\mathcal{Q}_{\textrm{ori},\infty}^\textrm{mag} (W^\textrm{mag})\geq 2$  for $B \kappa <0$, but does not hold for $B \kappa >0$.

We remark that the original TUR
is valid for all parameter regions when we take the Lorentz force as the irreversible one.
In this case, the EP rate includes an extra positive {\em unconventional} term $\sigma_\textrm{s}^\textrm{unc}=2B^2 (k+\kappa B/\gamma)/(m\gamma K)$~\cite{Kwon,LeeS}.
However, it is not clear at this moment that
this choice may guarantee the validity of the original TUR  in a more generalized model
with a magetic field as well as nonlinear nonconservative forces.

\subsection{Active matter}

We consider an overdamped `self-propelled' or `active' particle moving along a one-dimensional ring. The dynamics of an active particle is sometimes described by the active Ornstein-Uhlenbeck process (AOUP)~\cite{Fodor,Madal}, which is given by
\begin{align}\label{eq:x_am}
\dot{x} = -\mu \partial_x \Phi + f_\textrm{nc} +\eta,
\end{align}
where $\mu$ is the mobility, $\Phi$ is a potential applied to the particle,  $f_\textrm{nc}$ is an applied nonconservative force, and $\eta$ is a colored noise satisfying $\langle \eta(t) \eta(0)\rangle = (D/\tau ) e^{-|t|/\tau}$. In the AOUP description,
the evolution of $\eta$ is simply described by
\begin{align}\label{eq:tau_am}
\tau \dot{\eta} = -\eta + \sqrt{2D} \xi,
\end{align}
where $\tau$ is the persistence time and  $\xi$ is a white Gaussian noise satisfying $\langle \xi(t) \xi(t^\prime)\rangle = \delta (t-t^\prime)$. This overdamped motion with a colored noise can be written in the form of an  underdamped dynamics. By introducing the auxiliary velocity $v \equiv \dot{x}$ and mass $m \equiv \tau/\mu$, we have~\cite{Fodor,Madal}
\begin{align}
m\dot{v} = -\partial_x \Psi - \tau v \partial_x^2 \Phi +\gamma f_\textrm{nc} + m \dot{f}_\textrm{nc} - \gamma v +\sqrt{2 \gamma T} \xi, \label{eq:active_underdamped}
\end{align}
where $\Psi =\Phi +\tau \partial_t \Phi$ , $\gamma \equiv 1/\mu$,
and the auxiliary temperature by  $T \equiv D/\mu $.
Equation~\eqref{eq:active_underdamped} describes the underdamped motion of a Brownian particle with an external force $F=-\partial_x \Psi - \tau v \partial_x^2 \Phi +\gamma f_\textrm{nc} + m \dot{f}_\textrm{nc} $ in a reservoir with temperature $T$. As $F$ depends on velocity $v$, the effect of a velocity-dependent force should be considered for the uncertainty relation.

Here, we test our generalized TUR with a tilted periodic potential~\cite{Vu1,Hyeon}:
\begin{align}
f_\textrm{nc} = f~ (\textrm{constant})~~\textrm{and}~~~\Phi = \frac{AL}{2\pi n} \sin\left( \frac{2\pi n}{L}x \right) , \label{eq:activePotential}
\end{align}
where $A$ is the amplitude of the potential, $L$ is the length of the ring, and $n$ is an integer number. As there is no explicit time dependence in $f_\textrm{nc}$ and $\Phi$,
$\dot{f}_\textrm{nc} =0$ and $\Psi = \Phi$. We choose $F^\textrm{ir}$ as a velocity-dependent part of $F$. Thus, we have
\begin{align}
F^\textrm{rev} &= -A \cos \left(\frac{2\pi n}{L} x \right) +\gamma f,  \nonumber \\
F^\textrm{ir} &=  \frac{2 \pi n  m\mu A}{L}v \sin \left(\frac{2\pi n}{L} x \right), \nonumber \\
\mathcal{F} &= -2A \cos \left(\frac{2\pi n}{L} x \right) +2\gamma f -\frac{4 \pi n  m\mu A}{L}v \sin \left(\frac{2\pi n}{L} x \right).
\end{align}
The steady-state EP rate can be expressed as
\begin{align}
\sigma_\textrm{s} =  - \frac{A}{T}\left( \frac{2\pi n m \mu}{L} \right)^2  \left\langle v^3 \cos \left( \frac{2\pi n}{L} x \right) \right\rangle_\textrm{s} + \frac{\gamma f}{T} \langle v \rangle_\textrm{s}, \label{eq:activeEPrate}
\end{align}
of which the detailed derivation is presented in Appendix B.

It is not possible to proceed to calculate the TUR factor analytically with the periodic potential. For simplicity, we first consider the case of $\Phi=0$, where all calculations can be
done analytically, but there is no velocity-dependent force.  The auxiliary underdamped equation of motion,
Eq.~\eqref{eq:active_underdamped}, becomes
\begin{align}
\dot{x} = v,~~ m\dot{v} = \gamma f -\gamma v + \sqrt{2\gamma T} \xi. \label{eq:active_const_force}
\end{align}
Here, we take $F=\gamma f =F^\textrm{rev}$ and $F^\textrm{ir}=0$, thus $ \mathcal{F}=2F$. By making a change of variables as $V = v-f$ and $X = x -ft$, Eq.~\eqref{eq:active_const_force} becomes $\dot{X}=V$ and $m\dot{V} =-\gamma V + \sqrt{2\gamma T} \xi$, which
describes a Brownian motion  without an external force. Thus, it is straightforward to calculate $\langle v \rangle_\textrm{s} = f$, $\langle x \rangle_\textrm{s} = f \mathcal{T}$, $\langle v^2 \rangle_\textrm{s} = f^2 + T/m$, and $\langle x^2 \rangle_\textrm{s} = f^2 \mathcal{T}^2 + 2T \mathcal{T}/\gamma$ from the facts $\langle V \rangle_\textrm{s} = 0$, $\langle X \rangle_\textrm{s} = 0$, $\langle V^2 \rangle_\textrm{s} = T/m$, and $\langle X^2 \rangle_\textrm{s} = 2T \mathcal{T}/\gamma$. In addition, the EP rate is $\sigma_\textrm{s} = \gamma f \langle v \rangle_\textrm{s}/T = \gamma f^2 /T$ and the steady-state distribution function is $P^\textrm{ss} (v)= \sqrt{m/(2\pi T)} \exp[-m (v-f)^2 / (2T)]$. Using these results, $\Sigma_\textrm{uv}$ of the active matter becomes
\begin{align}
\Sigma_\textrm{uv}^\textrm{am} = \left[ \left( \frac{\gamma f^2}{T} + \frac{4\gamma}{m} \right) \mathcal{T} + \frac{2 m f^2}{T} +4 \right]/( 1+ \langle \Theta^\prime \rangle_\textrm{s} /\langle \Theta \rangle_\textrm{s})^2.
\end{align}

We consider the accumulated work $W^\textrm{mag}$ done by the constant force $f$, giving $\Lambda=f$ and
$\langle {W^\textrm{mag}}^\prime \rangle_\textrm{s}=0$.
It is easy to calculate its average and variance as
\begin{align}
\frac{\textrm{Var} [W^\textrm{am}] }{\langle W^\textrm{am} \rangle_\textrm{s}^2} = \frac{\textrm{Var} [x] }{\langle x\rangle_\textrm{s}^2} = \frac{2 T}{\gamma f^2 \mathcal{T}}.
\end{align}
Then, the original TUR for this active system gives
\begin{align}
\mathcal{Q}_\textrm{ori}^\textrm{am} (W^\textrm{am}) \equiv \frac{\textrm{Var} [W^\textrm{am}]}{ \langle W^\textrm{am} \rangle_\textrm{s}^2} \mathcal{T} \sigma_\textrm{s} = 2,
\end{align}
which always reaches the lowest bound of the original TUR.
On the other hand, our generalized TUR gives
\begin{align}\label{eq:Qam_withoutA}
\mathcal{Q}_\textrm{uv}^\textrm{am} (W^\textrm{am}) \equiv \frac{\textrm{Var} [W^\textrm{am}]}{ \langle W^\textrm{am} \rangle_\textrm{s}^2} \Sigma_\textrm{uv}^\textrm{am} = 2 + \frac{8T}{f^2 m} + \frac{4T}{\gamma f^2 \mathcal{T} } \left( \frac{mf^2}{T} +2 \right),
\end{align}
which is always larger than $2$. The lowest bound is reachable in the limit of
both $\mathcal{T}$ and $f\rightarrow\infty$.

For nonzero $\Phi$, we perform numerical simulations.
For convenience, we first reduce the number of parameters of the system by rescaling the units of length, time, and energy. To this end, we introduce the dimensionless variables
\begin{align}
    x_0 \equiv \frac{n}{L}x,~~
    t_0 \equiv \frac{t}{\tau},~~
    \eta_0 \equiv \frac{n \tau}{L} \eta,
\end{align}
and $\xi_0 (t_0) \equiv \sqrt{\tau} \xi (t)$
satisfying
$\langle \xi_0 (t_0) \xi_0(t_0^\prime) \rangle = \delta(t_0 - t_0^\prime)$.
Equations~\eqref{eq:x_am} and \eqref{eq:tau_am}
are then written in terms of the dimensionless variables and parameters
as
\begin{equation}
    \partial_{t_0} x_0 = - \mu_0 \partial_{x_0} \Phi_0
    + f_0 + \eta_0, \label{eq:active_reduced1}
\end{equation}
and
\begin{equation}
    \partial_{t_0} \eta_0 = - \eta_0 + \sqrt{2 \mu_0} \xi_0,
    \label{eq:active_reduced2}
\end{equation}
where $\mu_0 \equiv (n / L)^2 \tau T \mu$, $f_0 \equiv n \tau f / L$, and $\Phi_0 \equiv \Phi / T = (A_0 / 2 \pi) \sin (2 \pi x_0)$ with $A_0 = L A / (n T)$. Thus, the system is characterized by the three dimensionless parameters, $\mu_0$, $f_0$, and $A_0$. We numerically integrate the underdamped equations, Eqs.~\eqref{eq:active_reduced1} and \eqref{eq:active_reduced2}, using the second-order integrator~\cite{Ciccotti}.
Note that the boundary term $\Omega$ cannot be calculated without knowing the explicit form of the steady-state distribution $p^\textrm{ss}(x,v)$. Thus, we consider the infinite-time limit
($\mathcal{T}\rightarrow\infty$), where the boundary term can be neglected.  We then calculate $\langle \dot{W}^\textrm{am} \rangle_\textrm{s}$ and the diffusion coefficient $D_W^\textrm{am} = \lim_{\mathcal{T} \rightarrow \infty} \textrm{Var}[W^\textrm{am}]/(2\mathcal{T})$ via extrapolation from $10^5$ different trajectories.

\begin{figure}
\centering
\includegraphics[width=\linewidth]{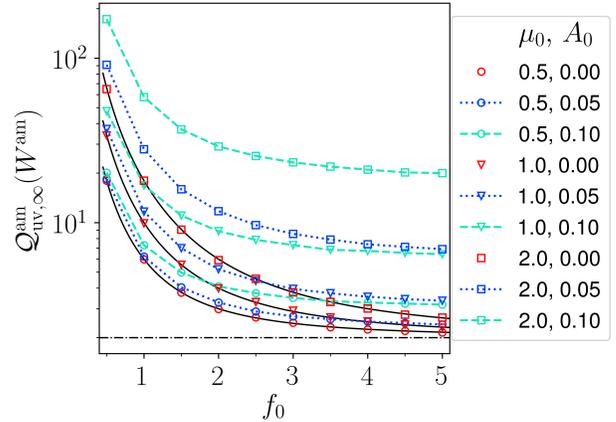}
\caption{
The TUR factor $\mathcal{Q}$  as a function of the dimensionless force $f_0$ in the
infinite-time limit for the active matter system for various $\mu_0$ and $A_0$.
Circles, triangles, and squares denote the numerical results for $\mu_0 = 0.5$, $1$, and $2$, respectively.
Symbols with no line (red), dotted line (blue), and dashed line (cyan) represent $A_0 = 0$, $A_0=0.05$, and $A_0 = 0.1$, respectively. The analytic results~\eqref{eq:Qam_withoutA} for $A = 0$ are drawn by the black solid curves. The black dashed-dotted line indicates the bound $\mathcal{Q}=2$.
} \label{fig:active}
\end{figure}

The results are presented in Fig.~\ref{fig:active}. As shown in the figure, $\mathcal{Q}_{\textrm{uv},\infty}^\textrm{am} (W^\textrm{am})$ satisfies the bound~\eqref{eq:main_ineq}.
When $A_0 = 0$,
the numerical data are consistent with our analytic result \eqref{eq:Qam_withoutA}. For all parameter spaces, it seems that
$\mathcal{Q}_{\textrm{uv},\infty}^\textrm{am} (W^\textrm{am})$ is a monotonically decreasing function of $f_0$ and a monotonically increasing function of $\mu_0$ and $A_0$.
Furthermore, as the effect of
$A_0$ becomes negligible for large $f$~\cite{Hyeon}, it is expected that all curves can be fitted by Eq.~\eqref{eq:Qam_withoutA}
and saturate to the bound $2$ in the large $f$ limit.

\section {Conclusion and Discussion}

In summary, we derived a generalized TUR for underdamped Langevin dynamics with a velocity-dependent force by extending the theory of Vu and Hasegawa~\cite{Vu1}, and showed that a velocity-dependent force only changes the form of the dynamical activity function, compared to the case without a velocity-dependent force. In particular, our bound does not depend on the choice of reversible and irreversible forces, in sharp contrast to the original TUR~\cite{Barato} and the modified TUR~\cite{Vu1}, where the bounds are dependent on the choice of reversible force in the presence of velocity-dependent forces.

We examine our bound in the context of three important systems pertaining to a velocity-dependent force: a molecular refrigerator, Brownian motion in a magnetic field, and active matter. In each case, we calculate the original and our TUR for the work fluctuations and show that our TUR always holds, whereas the previously known TURs do not.

In the original TUR, the equality when the TUR factor reaches the bound is attained when the distribution is Gaussian~\cite{Hyeon}. Thus, the lowest bound is reached on approaching the reversible limit. However, in the case of our generalized TUR, the equality is not simply attained on reaching the reversible limit. Instead, certain system-dependent specific limits are necessary to obtain the lowest bound. Moreover, it seems that the lowest bound is not achievable for some systems, e.g., a Brownian particle in a magnetic field. Thus, in these cases, our bound
becomes loose.

Finally, we comment on the uncertainty relation derived from the fluctuation theorems. Hasegawa and Vu~\cite{Vu3} recently presented another type of TUR, as follows:
\begin{align}
\frac{\textrm{Var} [\Theta]}{ \langle \Theta \rangle^2} \geq \frac{2}{e^{\Delta S_\textrm{tot}}-1}, \label{eq:GTUR}
\end{align}
where they assume that the detailed fluctuation theorem is satisfied for $\Delta S_\textrm{tot}$ (EP during the time interval $\mathcal{T}$), and the current $\Theta$ is antisymmetric under time-reversal operation of a trajectory, that is, $\Theta[\Gamma] = -\Theta[\Gamma^\dagger]$. This TUR is known to be valid for time-discrete Markov chains and time-dependent driving with a periodic and time-symmetric protocol~\cite{Proesmans,Vu3}. In the molecular-refrigerator case, the work current is symmetric under time reversal, such that the above TUR cannot be applied. In addition, the above TUR provides an exponentially loose bound for the long-time limit, whereas our TUR constrains the relative fluctuations linearly.

\begin{acknowledgments}
This research was supported by the NRF Grant No.~2017R1D1A1B06035497 (HP).
\end{acknowledgments}

\appendix

\section{Variation of work in the molecular refrigerator}

The velocity $v$ and work $W^\textrm{mr}$
are stochastic random variables.
From the equation of motion \eqref{eq:MolRef} and
the definition of work,
one can write the stochastic differential equation for the velocity and work as
\begin{align}
    \dot{v} &= -\frac{\alpha + \gamma}{m} v + \frac{1}{m} \xi, \\
    \dot{W}^\text{mr} &= - \alpha v^2. \label{eqA:work_equation}
\end{align}
For simplicity, we will omit the superscript `$\text{mr}$' for the following discussion
in this section.

By using $dW^2/dt = 2W\dot{W}$ and $d(v^2W)/dt = 2v \circ \dot{v}W+v^2 \dot{W}$, we have the following equations:
\begin{equation}\label{d_eq:W2}
    \partial_t \langle W^2 \rangle_\textrm{s}
    = - 2\alpha \langle v^2 W \rangle_\textrm{s}
\end{equation}
and
\begin{align}\label{d_eq:v2W}
    \partial_t \langle v^2 W \rangle_\textrm{s}
    = - 2 \frac{\alpha + \gamma}{m}
    \langle v^2 W \rangle_\textrm{s}
    - \left ( \frac{\alpha T^\prime}{m} \right)^2
    \left ( \frac{3}{\alpha}
    + \frac{2(\alpha+\gamma)}{m \alpha} t \right ).
\end{align}
In the derivation of Eq.~\eqref{d_eq:v2W},  we used
$\langle v^2 \rangle_\textrm{s} = T^\prime / m$
and the property of the Gaussian distribution,
$\langle v^4 \rangle_\textrm{s} =
3 \langle v^2 \rangle_\textrm{s}^2  $.
By solving  \eqref{d_eq:v2W}, we get
\begin{equation}
    \langle v^2 W \rangle_\textrm{s}
    = - \frac{\alpha (T^\prime)^2}{m (\alpha + \gamma)}
    \left [ 1 - \exp(- 2(\alpha+\gamma)t/m) \right ]
    - \alpha \left (\frac{T^\prime}{m} \right)^2 t. \label{eqA:sln}
\end{equation}
Plugging the solution~\eqref{eqA:sln} into Eq.~\eqref{d_eq:W2}
and integrating up to $\mathcal{T}$ yields
\begin{equation}
    \langle W^2 \rangle_\textrm{s}
    = 2 \frac{(\alpha T^\prime)^2}{m (\alpha + \gamma)}
    \left ( \mathcal{T} - \frac{m[1 - \exp(- 2(\alpha+\gamma)\mathcal{T}/m)]}
    {2(\alpha+\gamma)} \right )
    + \langle W \rangle_\textrm{s}^2,
\end{equation}
which leads to the variance of work, Eq.~\eqref{eq:MolRef_Wvar}.

\section{The EP rate in the active matter with a tilted periodic potential}

From the Onsager-Machlup formalism~\cite{Onsager}, the probability densities for the forward $\Gamma$ and time-reversal $\tilde{\Gamma}$ trajectories of Eq.~\eqref{eq:active_underdamped} in the Stratonovich convention are given by
\begin{align}
\mathcal{P}[\Gamma] &= \mathcal{N} P (\textit{\textbf{x}}_0, \textit{\textbf{v}}_0)  \exp \left[ -\frac{\mu}{4T} \int_0^\mathcal{T} dt~ G^2 - \frac{1}{2} \int_0^\mathcal{T} dt ~ g      \right], \nonumber \\
\mathcal{P}[\tilde{\Gamma}] &= \mathcal{N} P (\textit{\textbf{x}}_\mathcal{T}, \textit{\textbf{v}}_\mathcal{T})  \exp \left[ -\frac{\mu}{4T} \int_0^\mathcal{T} dt ~\tilde{G}^2 - \frac{1}{2} \int_0^\mathcal{T} dt  ~\tilde{g}    \right],
\end{align}
respectively, where $\mathcal{N}$ is a normalization constant and $G$, $\tilde{G}$, $g$, $\tilde{g}$ are give by
\begin{align}
G &\equiv  m \dot{v} + \partial_x \Phi + \tau v\partial_x^2 \Phi -\gamma f +\gamma v,  \nonumber \\
\tilde{G} &\equiv m \dot{v} + \partial_x \Phi + \tau (-v)\partial_x^2 \Phi -\gamma f + \gamma (-v), \nonumber \\
g &\equiv -\mu \partial_v  v \partial_x^2 \Phi -1/\tau,  \nonumber \\
\tilde{g} &\equiv -\mu \partial_{(-v)}  (-v) \partial_x^2 \Phi -1/\tau.
\end{align}
Thus, the total EP can be written as~\cite{Udo_review, Fodor}
\begin{align}
\Delta S_\textrm{tot} [\Gamma]
= \ln \frac{\mathcal{P}[\Gamma]}{ \mathcal{P}[\tilde{\Gamma}] }
= \ln \frac{P (\textit{\textbf{x}}_0, \textit{\textbf{v}}_0)}{ P (\textit{\textbf{x}}_\mathcal{T}, \textit{\textbf{v}}_\mathcal{T})}
-\frac{\mu}{T} \int_0^\mathcal{T} dt ~ \mathcal{G},
\end{align}
where $\mathcal{G} \equiv \tau m (\dot{v} \partial_x )( v \partial_x ) \Phi + \gamma m v \circ \dot{v} + \tau \partial_x \Phi (v \partial_x)\partial_x \Phi + \gamma (v \partial_x )\Phi - \gamma \tau f (v \partial_x) \partial_x \Phi - \gamma^2 f v$.
As $\Phi$ has only $x$ dependence, we can use the relation $v \partial_x = d/dt$. Then, we can show that
\begin{align}
& \int_0^\mathcal{T} dt~ v\circ \dot{v} = \frac{1}{2} \left(v_\mathcal{T}^2 - v_0^2 \right), \label{eq:integration1}
\end{align}
\begin{align}
& \int_0^\mathcal{T} dt~ v \partial_x \Phi = \Phi_\mathcal{T} - \Phi_0,  \label{eq:integration2}
\end{align}
\begin{align}
& \int_0^\mathcal{T} dt~ \partial_x \Phi (v\partial_x) \partial_x \Phi = \frac{1}{2} \left( (\partial_x \Phi_\mathcal{T})^2 - (\partial_x \Phi_0)^2 \right), \label{eq:integration3}
\end{align}
\begin{align}
& \int_0^\mathcal{T} dt~ (v \partial_x) \partial_x \Phi = \partial_x \Phi_\mathcal{T} - \partial_x \Phi_0,  \label{eq:integration4}
\end{align}
\begin{align}
& \int_0^\mathcal{T} dt~ (\dot{v} \partial_x)(v \partial_x) \Phi = -\int_0^\mathcal{T} (v\partial_x)^3 \Phi, \label{eq:integration5}
\end{align}
\begin{align}
& \int_0^\mathcal{T} dt~ v  = x_\mathcal{T} - x_0. \label{eq:integration6}
\end{align}
As Eqs.~\eqref{eq:integration1}, \eqref{eq:integration2}, \eqref{eq:integration3}, and \eqref{eq:integration4} are boundary terms, their steady-state averages vanish. Thus, only remaining terms in the steady state are Eqs.~\eqref{eq:integration5} and \eqref{eq:integration6}. Then, the steady-state average of $\Delta S_\textrm{tot}$ becomes
\begin{align}
\langle \Delta S_\textrm{tot} \rangle_\textrm{s} = \mathcal{T}  \sigma_\textrm{s} = \mathcal{T} \left[ \frac{m^2 \mu^2}{T} \langle v^3 \partial_x^3 \Phi \rangle_\textrm{s} +\frac{\gamma f}{T} \langle v \rangle_\textrm{s} \right].
\end{align}

\vfil\eject

\end{document}